# Maximal Atom-Photon Entanglement in a Double-Λ Quantum System

**Zeinab Kordi, Saeed Ghanbari and Mohammad Mahmoudi[†]**

Department of Physics, University of Zanjan, University Blvd, 45371-38791,

Zanjan, Iran

[†]Corresponding author's email: mahmoudi@znu.ac.ir

Tel: +98-24-33052521

Fax: +98-24-33052264

Abstract. The atom-photon entanglement of the dressed atom and its spontaneous emission in a Double-Λ closed-loop atomic system is studied in multi-photon resonance condition. It is shown that, even in the absence of quantum interference due to the spontaneous emission, the von Neumann entropy is phase-sensitive and it can be controlled by either intensity or relative phase of the applied fields. It is demonstrated that, for the special case of Rabi frequency of the applied fields the system is maximally entangled. Moreover, the open-loop configuration is considered and it is shown that the degree of entanglement (DEM) can be controlled by intensity of the applied fields. Furthermore, in electromagnetically induced transparency condition, the system is disentangled. Such a system can be used for quantum information processing via entanglement using optical switching.

## 1. Introduction

Quantum entanglement is one of the most fantastic and mysterious features of quantum mechanics which has no counterpart in classical physics. In a 1935 paper by Einstein, Podolsky and Rosen [1], for first the time, this wonderful phenomenon was presented. The term "Entanglement" was coined by Schrodinger in his response to Einstein's letter [2] in 1935 on the foundations of quantum mechanics. Generally, the entangled state of a quantum system is not defined as the tensor product of the quantum states of the subsystems [3]. In this case, a measurement on one of them provides information on the other ones.

In fact, entanglement is important for a host of applications and has been widely recognized as an important tool in many quantum communication protocols, quantum teleportation and entanglement swapping [4], quantum super dense coding [5], quantum error correction [6, 7], quantum cryptography [8-10] entanglement distillation [11] and quantum computing [12].

Entanglement can occur because of the interaction between different parts of a system consisting of atoms, photons or a mixture of atoms and photons which can be measured in different methods. Reduced quantum entropy is a good measure which can be used to quantify entanglement for bipartite systems [13].

Atom-photon entanglement due to the interaction between the matter and light can play a key role in quantum information storage. Entanglement between atom and its spontaneous emission field in a $\Lambda$ has been reported [14]. The effect of spontaneously generated coherence on the atom–photon entanglement has been also investigated [15]. Moreover, it has been demonstrated that, in the presence of quantum interference induced by spontaneous emission, the atom-photon entanglement is phase-sensitive [16]. Furthermore, it has been shown that the atom–photon entanglement in closed-loop three-level quantum systems can be controlled by relative phase of the applied fields [17].

On the other hand, the optical properties of double-$\Lambda$ quantum system have been extensively investigated [18-19]. The phase-sensitive electromagnetically induced transparency [20] and efficient nonlinear frequency conversion [21] have been investigated experimentally in this system. The light propagation through closed-loop atomic media beyond the multi-photon resonance condition has been also studied [22]. The double-lambda schemes have been used in the past to generate the non-classical states of light [23] and the entangled beams [24-26] from four-wave mixing. Recently, the coherent control of quantum entropy via quantum interference has been proposed [27]. More recently, it has been used to generate a single-photon frequency-bin entangled state [28] and all-optical transistor at ultralow light level which especially attractive for its potential applications in the quantum information field [29].

In this paper, we are interested in studying the dynamical behavior of entanglement of the atom and its spontaneous emission in a double-$\Lambda$ quantum system in multi-photon resonance condition. We show that, even in the absence of quantum interference due to the spontaneous emission, the atom is entangled with its spontaneous emission. It is demonstrated that the steady state DEM can be controlled by either intensity or relative phase of the applied fields. By considering appropriate values for the optical laser fields, the maximal DEM is obtained. Furthermore, the atom-photon entanglement disappears in the electromagnetically induced transparency condition.

## 2. Models and equations

We consider a four-level double-Λ quantum system in the closed-loop configuration as illustrated in Fig. 1. This set contains two metastable lower states $|1\rangle$ and $|2\rangle$ as well as two excited states $|3\rangle$ and $|4\rangle$. The transitions $|1\rangle$-$|3\rangle$, $|2\rangle$-$|3\rangle$, $|1\rangle$-$|4\rangle$ and $|2\rangle$-$|4\rangle$ are excited by four coherent laser fields. The spontaneous emission rates from level $|i\rangle$ ($i \in \{3,4\}$) to the levels $|j\rangle$ ($j \in \{1,2\}$) are denoted by $2\gamma_{ij}$. As a realistic example, we consider Rubidium atoms in a vapor cell [30, 31]. The laser coupling of the transition $|i\rangle \rightarrow |j\rangle$ is characterized by the frequency $\omega_{ij}$ and the wave vector $k_{ij}$. So, the electromagnetic driving fields are

$$E_{ij} = E_{ij}\hat{n}_{ij}e^{-i(\omega_{ij}t - \vec{k}_{ij}\cdot\vec{r} + \phi_{ij})} + c.c., \tag{1}$$

where $\hat{n}_{ij}$ and $\phi_{ij}$ are the polarization unit vector and the absolute phase, respectively. The semi-classical Hamiltonian in the rotating wave and dipole approximation is given by [32, 33]

$$H = \sum_{j=1}^{4} E_j |j\rangle\langle j| - \sum_{l=3}^{4}\sum_{m=1}^{2} \hbar g_{lm} e^{-i\alpha_{lm}} |l\rangle\langle m| + \text{H.C}, \tag{2}$$

where $g_{ij} = E_{ij}(\hat{e}_{ij}\cdot d_{ij})/\hbar$ is the Rabi frequency with $d_{ij}$ as the atomic dipole moment of the corresponding transition. Also, $E_j$ ($j \in \{1,...,4\}$) denotes the energy of the involved states. The transition frequencies and laser field detuning are defined by $\bar{\omega}_{ij} = (E_i - E_j)/\hbar$ and $\Delta_{ij} = \omega_{ij} - \bar{\omega}_{ij}$, respectively. The exponents are given by $\alpha_{ij} = \omega_{ij}t - \vec{\kappa}_{ij}\vec{r} + \varphi_{ij}$.

By changing the reference frame and using the interaction picture, the Hamiltonian can be derived as [22]

$$H = \hbar(\Delta_{32} - \Delta_{31})\tilde{\rho}_{22} - \hbar\Delta_{31}\tilde{\rho}_{33} + \hbar(\Delta_{32} - \Delta_{31} - \Delta_{42})\tilde{\rho}_{44} - \\ \hbar(g_{31}\tilde{\rho}_{31} + g_{32}\tilde{\rho}_{32} + g_{42}\tilde{\rho}_{42} + g_{41}\tilde{\rho}_{41}e^{-i\Phi} + \text{H.C}), \tag{3}$$

where $\Phi = \Delta t - \vec{K}\vec{r} + \varphi_0$, $\Delta = (\Delta_{32} + \Delta_{41}) - (\Delta_{31} + \Delta_{42})$, $\vec{K} = (\vec{\kappa}_{32} + \vec{\kappa}_{41}) - (\vec{\kappa}_{31} + \vec{\kappa}_{42})$, $\varphi_0 = (\varphi_{32} + \varphi_{41}) - (\varphi_{31} + \varphi_{42})$.

The density matrix elements in the rotating frame and rotating wave approximation are then derived using Liouville's theorem:

$$\frac{\partial}{\partial t}\rho_{11} = ig_{31}^*\rho_{31} - ig_{31}\rho_{13} + ig_{41}^*\rho_{41}e^{i\Phi} - ig_{41}\rho_{14}e^{-i\Phi} + 2\gamma_{13}\rho_{33} + 2\gamma_{14}\rho_{44},$$

$$\frac{\partial}{\partial t}\rho_{22} = ig_{32}^*\rho_{32} - ig_{32}\rho_{23} + ig_{42}^*\rho_{42} - ig_{42}\rho_{24} + 2\gamma_{23}\rho_{33} + 2\gamma_{24}\rho_{44},$$

$$\frac{\partial}{\partial t}\rho_{33} = -ig_{31}^*\rho_{31} - ig_{32}^*\rho_{32} + ig_{31}\rho_{13} + ig_{32}\rho_{23} - 2\gamma_3\rho_{33},$$

$$\frac{\partial}{\partial t}\rho_{12} = i(\Delta_{32} - \Delta_{31})\rho_{12} + ig_{31}^*\rho_{32} - ig_{32}\rho_{13} + ig_{41}^*\rho_{42}e^{i\Phi} - ig_{42}\rho_{14} - \Gamma_{12}\rho_{12},$$

$$\frac{\partial}{\partial t}\rho_{13} = -i\Delta_{31}\rho_{13} + ig_{31}^*(\rho_{33} - \rho_{11}) - ig_{32}^*\rho_{12} + ig_{41}^*\rho_{43}e^{i\Phi} - \Gamma_{13}\rho_{13},$$

$$\frac{\partial}{\partial t}\rho_{14} = i(\Delta_{32} - \Delta_{31} - \Delta_{42})\rho_{14} + ig_{41}^*e^{i\Phi}(\rho_{44} - \rho_{11}) - ig_{42}^*\rho_{12} + ig_{31}^*\rho_{34} - \Gamma_{14}\rho_{14},$$

$$\frac{\partial}{\partial t}\rho_{23} = -i\Delta_{32}\rho_{23} + ig_{32}^*(\rho_{33} - \rho_{22}) - ig_{31}^*\rho_{21} + ig_{42}^*\rho_{43} - \Gamma_{23}\rho_{23},$$

$$\frac{\partial}{\partial t}\rho_{24} = -i\Delta_{42}\rho_{24} + ig_{42}^*(\rho_{44} - \rho_{22}) - ig_{41}^*\rho_{21}e^{i\Phi} + ig_{32}^*\rho_{34} - \Gamma_{24}\rho_{24},$$

$$\frac{\partial}{\partial t}\rho_{34} = -i(\Delta_{42} - \Delta_{32})\rho_{34} + ig_{31}\rho_{14} + ig_{32}\rho_{24} - ig_{41}^*\rho_{31}e^{i\Phi} - ig_{42}^*\rho_{32} - \Gamma_{34}\rho_{34},$$

$$\rho_{44} = 1 - \rho_{11} - \rho_{22} - \rho_{33}, \tag{4}$$

where $\gamma_j = \gamma_{1j} + \gamma_{2j}$. Parameter $\Gamma_{ij} = (2\gamma_i + 2\gamma_j)/2$ is the damping rate of the $|i\rangle \to |j\rangle$ transition coherence. For simplicity, the spontaneous emission rates of the excited levels are assumed to be equal. To work out these equations, the phase matching ($\vec{K} = 0$) and multi-photon resonance ($\Delta = 0$) conditions should be satisfied by the applied fields.

## 3. The Evolution of Entropy and Atom-photon Entanglement

Now, we are interested in calculating the degree of entanglement via reduced entropy. Although other entanglement measures exist, reduced entropy is one of the important tools which can be used to quantify it. [34].

The system is entangled if it is not separable. Mathematically, the bipartite quantum system is called separable, when its density operator can be written as [35]

$$\rho_{AB} = \rho_A \otimes \rho_B. \tag{5}$$

We consider the atom and vacuum field initially in a disentangled pure state which means that all of the atoms are initially in just one level. If the overall system is pure, the entropy of one subsystem can be used to measure its degree of entanglement with other subsystems. For bipartite pure states, the von Neumann entropy of the reduced states is a good measure of DEM [36-37]. The quantum mechanical von Neumann entropy can be defined by

$$S(\rho) = -Tr\rho \ln \rho ,  \qquad (6)$$

where $\rho$ stands for the density operator of the system.

The reduced density operator of the atoms (spontaneous emission) as the first (the second) subsystem is defined by:

$$\rho_{A(F)} = Tr_{F(A)}\{\rho_{AF}\} ,  \qquad (7)$$

where $\rho_{AF}$ is density operator of the pure state for two subsystems $A$ and $F$. Therefore, the partial von Neumann entropy corresponding to the reduced density operator is derived as [13, 3]:

$$S_{A(F)} = -Tr(\rho_{A(F)} \ln \rho_{A(F)}). \qquad (8)$$

Araki and Lieb have shown that for a bipartite quantum system composed of two subsystems A and F (say the atom and field) at any time t, the system and subsystem entropies satisfy an inequality as bellow [36- 38]:

$$|S_A(t) - S_F(t)| \leq S_{AF}(t) \leq |S_A(t) + S_F(t)|, \qquad (9)$$

where $S_{AF}$ is the total entropy of the composite system.

Based on Equation (9), for a closed atom–field system in which both of them start from a pure state, the entropies of two interacting subsystems will be precisely equal at all times after the interaction of the two subsystems is switched on. The DEM for atom-field entanglement is defined by:

$$DEM(t) = S_A = S_F = -(\sum_{j=1}^{4} \lambda_j \ln \lambda_j), \qquad (10)$$

where $\lambda_j$ denote the eigenvalues of the reduced density matrix.

In fact, the generated light after interaction with atomic system has some information about atomic properties which can be directly measured by quantum discord [39-40]. The atom-photon entanglement behavior can be understood via population distribution of the dressed states.

## 4. Results and discussions

Steady state behavior of the density matrix equations of motion are numerically investigated in multi-photon resonance condition. In all of the calculations, for simplicity, it is supposed that $\hbar = 1$. All of the parameters in computer codes are reduced to dimensionless units and are scaled as $\gamma_{13} = \gamma_{23} = \gamma_{14} = \gamma_{24} = \gamma$, $\gamma = 1$. All plots are sketched in the unit of $\gamma$.

Now, we are interested in studying the dynamical behavior of entanglement for different values of parameters. Figure 2 shows the time evolution of DEM in both closed-loop for $\varphi_0 = 0$ (solid), $\pi$ (dashed) and open-loop (dash-dotted) configurations. The open-loop is created by removing one of the optical driving fields ($g_{32} = 0$). Parameters used are $\gamma_{13} = \gamma_{23} = \gamma_{14} = \gamma_{24} = \gamma = 1$, $\gamma_3 = \gamma_{13} + \gamma_{23}$, $\Gamma_{13} = \gamma_{13} + \gamma_{23}$, $\Gamma_{23} = \gamma_{13} + \gamma_{23}$, $\Gamma_{24} = \gamma_{14} + \gamma_{24}$, $\Gamma_{14} = \gamma_{14} + \gamma_{24}$, $\Gamma_{34} = \gamma_{13} + \gamma_{23} + \gamma_{14} + \gamma_{24}$, $\Gamma_{12} = \gamma$, $\Delta_{42} = \Delta_{32} = \Delta_{31} = 0$, $g_{31} = 3\gamma$, $g_{42} = 3\gamma$, $g_{41} = 3\gamma$, $g_{32} = 3\gamma$ (solid and dashed) and $g_{32} = 0$ (dash-dotted). An investigation on Fig. 2 shows that, the steady state entanglement can occur for both of these configurations. Also, even in the absence of quantum interference due to the spontaneous emission, the steady state DEM in the closed-loop configuration depends on the relative phase of the applied fields. It is worth noting that, in open–loop scheme the DEM is not vanished.

In Fig.3, we illustrate DEM versus relative phase of the applied fields for double-$\Lambda$ (solid) and open-loop (dashed) configurations. Parameters are $g_{31} = 3\gamma$, $g_{42} = 3\gamma$, $g_{41} = 3\gamma$, $g_{32} = 3\gamma$ (solid) and $g_{32} = 0$ (dashed). Other parameters are same as in Fig. 2. It is clearly seen that, the DEM for the open-loop system does not depend on the relative phase of the applied fields, but it does on the phase for double-$\Lambda$ configuration [41]. Thus, relative phase can play a major role in controlling the entanglement and the behavior of entanglement.

The dynamical behavior of DEM in phase switching for closed-loop and intensity switching for open-loop configurations is shown in Fig. 4. The parameters are same as in Fig. 2. DEM starts from zero at $t = 0$ and increases during interaction. It is realized that, DEM can be controlled by either relative phase or intensity of the applied fields.

This may provide a possibility for quantum information processing via entanglement using optical switching [42].

Let us focus more on the effect of applied fields on the DEM in the open-loop configuration. In this system, the DEM can be controlled by intensity of the applied fields. In Fig. 5, we plot the behavior of the steady state DEM versus Rabi frequencies $g_{31}$ and $g_{41}$. Here again, the parameters are same as in Fig. 2. It is clear that, DEM increases by increasing the intensity of the driving fields. Note that, the system is entangled even for $g_{31}=g_{32}=0$, but disentanglement is resulted for small values of $g_{41}$ or $g_{42}$ because of establishing the electromagnetically induced transparency [43]. In this case, all of atoms are populated in a single dark state.

We now introduce the dressed states generated by applied fields which are useful for understanding the optical properties of the system. The physics of the phenomenon can be explained via the population distribution of the dressed states. The dressed states for closed-loop configuration under the conditions,

$$g_{31}/g_{32} = g_{41}/g_{42},\ \varphi_0 = 2n\pi\ (n=0,1,2,...)\ ;\ \Delta_{32} - \Delta_{31} = \Delta_{42} - \Delta_{41} = 0,$$

can be written as [41]:

$$|D1\rangle = \frac{g_{31}/g_{32}}{\sqrt{1+g_{32}^2/g_{31}^2}}|1\rangle + \frac{1}{\sqrt{1+g_{32}^2/g_{31}^2}}|2\rangle,$$

$$|D2\rangle = \frac{g_{32}/g_{31}}{\sqrt{1+g_{32}^2/g_{31}^2}}|1\rangle - \frac{1}{\sqrt{1+g_{32}^2/g_{31}^2}}|2\rangle,$$

$$|D3\rangle = \frac{g_{32}/g_{42}}{\sqrt{1+g_{42}^2/g_{32}^2}}|3\rangle + \frac{1}{\sqrt{1+g_{42}^2/g_{32}^2}}|4\rangle,$$

$$|D4\rangle = \frac{g_{42}/g_{32}}{\sqrt{1+g_{42}^2/g_{32}^2}}|3\rangle - \frac{1}{\sqrt{1+g_{42}^2/g_{32}^2}}|4\rangle.$$

(11)

The corresponding populations can be calculated as follows:

$$\rho_{D1D1} = \frac{g_{32}^2/g_{42}^2}{1+g_{42}^2/g_{32}^2}\rho_{33} + \frac{g_{32}/g_{42}}{1+g_{42}^2/g_{32}^2}\rho_{34} + \frac{g_{32}/g_{42}}{1+g_{42}^2/g_{32}^2}\rho_{43} + \frac{1}{1+g_{42}^2/g_{32}^2}\rho_{44},$$

$$\rho_{D2D2} = \frac{g_{42}^2/g_{32}^2}{1+g_{42}^2/g_{32}^2}\rho_{33} - \frac{g_{42}/g_{32}}{1+g_{42}^2/g_{32}^2}\rho_{34} - \frac{g_{42}/g_{32}}{1+g_{42}^2/g_{32}^2}\rho_{43} + \frac{1}{1+g_{42}^2/g_{32}^2}\rho_{44},$$

$$\rho_{D3D3} = \frac{g_{31}^2/g_{32}^2}{1+g_{32}^2/g_{31}^2}\rho_{11} + \frac{g_{31}/g_{32}}{1+g_{32}^2/g_{31}^2}\rho_{12} + \frac{g_{31}/g_{32}}{1+g_{32}^2/g_{31}^2}\rho_{21} + \frac{1}{1+g_{32}^2/g_{31}^2}\rho_{22},$$

$$\rho_{D4D4} = \frac{g_{32}^2/g_{31}^2}{1+g_{32}^2/g_{31}^2}\rho_{11} - \frac{g_{32}/g_{31}}{1+g_{32}^2/g_{31}^2}\rho_{12} - \frac{g_{32}/g_{31}}{1+g_{32}^2/g_{31}^2}\rho_{21} + \frac{1}{1+g_{32}^2/g_{31}^2}\rho_{22}.$$ (12)

The dressed states of the open-loop configuration ($g_{32}=0$) for $g_{31}=g_{41}=g_{42}$ are given by

$$|d_1\rangle = \frac{1}{\sqrt{5-\sqrt{5}}}|1\rangle - \frac{1-\sqrt{5}}{2\sqrt{5-\sqrt{5}}}|2\rangle - \frac{1-\sqrt{5}}{2\sqrt{5-\sqrt{5}}}|3\rangle + \frac{1}{\sqrt{5-\sqrt{5}}}|4\rangle,$$

$$|d_2\rangle = \frac{-1}{\sqrt{5-\sqrt{5}}}|1\rangle + \frac{1-\sqrt{5}}{2\sqrt{5-\sqrt{5}}}|2\rangle - \frac{1-\sqrt{5}}{2\sqrt{5-\sqrt{5}}}|3\rangle + \frac{1}{\sqrt{5-\sqrt{5}}}|4\rangle,$$

$$|d_3\rangle = \frac{-1}{\sqrt{5+\sqrt{5}}}|1\rangle + \frac{1+\sqrt{5}}{2\sqrt{5+\sqrt{5}}}|2\rangle - \frac{1+\sqrt{5}}{2\sqrt{5+\sqrt{5}}}|3\rangle + \frac{1}{\sqrt{5+\sqrt{5}}}|4\rangle,$$

$$|d_4\rangle = \frac{1}{\sqrt{5+\sqrt{5}}}|1\rangle - \frac{1+\sqrt{5}}{2\sqrt{5+\sqrt{5}}}|2\rangle - \frac{1+\sqrt{5}}{2\sqrt{5+\sqrt{5}}}|3\rangle + \frac{1}{\sqrt{5+\sqrt{5}}}|4\rangle.$$ (13)

The population distribution in dressed states has a major role in determination of DEM.

The dynamical behavior of different dressed states population in both closed-loop for $\varphi_0=0$ (a), $\pi$ (b) and open-loop (c) configurations are displayed in Fig. 6. Parameters used are same as in Fig. 2. An investigation on Fig. 6 shows that for $\varphi_0=0$ the atoms leave one of the dressed states and population is distributed over three dressed states while for $\varphi_0=\pi$, it is distributed over four dressed states. Maximally entangled quantum states can be obtained for equal populations of the dressed states. Similar discussion is also valid for open-loop configuration.

**5. Conclusions**

In conclusion, we investigated the atom-photon entanglement in a four-level double-Λ closed-loop atomic system via the von Neumann entropy. The results were obtained in multi-photon resonance condition. It was found that, even in the absence of quantum interference due to the spontaneous emission, the DEM is phase-sensitive and nonzero steady state entropy was obtained for different values of relative phase of the applied fields. It was shown that the maximal DEM is obtained by choosing the suitable Rabi frequencies. DEM was also calculated for open-loop configuration and it was demonstrated that atom-photon entanglement can be controlled by intensity of the applied fields

**Figures captions**

**Figure 1.** Schematic diagram of the closed-loop four-level double-Λ quantum system. This system is driven by four optical laser fields. The spontaneous decays are denoted by the wiggly green lines.

**Figure 2.** Time evolution of DEM in both closed-loop for $\varphi_0 = 0$ (solid), $\pi$ (dashed) and non-closed loop (dash-dotted) configurations. The open-loop is created by removing one of the optical driving fields ($g_{32}=0$). The parameters are $\gamma_{13}=\gamma_{23}=\gamma_{14}=\gamma_{24}=\gamma=1$, $\gamma_3=\gamma_{13}+\gamma_{23}$, $\Gamma_{13}=\gamma_{13}+\gamma_{23}$, $\Gamma_{23}=\gamma_{13}+\gamma_{23}$, $\Gamma_{24}=\gamma_{14}+\gamma_{24}$, $\Gamma_{14}=\gamma_{14}+\gamma_{24}$, $\Gamma_{34}=\gamma_{13}+\gamma_{23}+\gamma_{14}+\gamma_{24}$, $\Gamma_{12}=\gamma$, $\Delta_{42}=\Delta_{32}=\Delta_{31}=0$, $g_{31}=3\gamma, g_{42}=3\gamma$, $g_{41}=3\gamma$, $g_{32}=3\gamma$ (solid and dashed), $g_{32}=0$ (dash-dotted).

**Figure 3.** DEM versus relative phase of the applied fields for double-Λ (solid) and open-loop (dashed) configurations. The parameters used are $g_{31}=3\gamma, g_{42}=3\gamma$, $g_{41}=3\gamma$, $g_{32}=3\gamma$ (solid) and $g_{32}=0$ (dashed). Other parameters are same as in Fig. 2.

**Figure 4.** The dynamical behavior of DEM in phase switching for closed-loop and intensity switching for open-loop configuration. The parameters are same as in Fig. 2.

**Figure 5.** Steady state behavior of DEM versus Rabi frequencies $g_{31}$ and $g_{41}$. The parameters used are same as in Fig. 2.

**Figure 6.** The dynamical behavior of different dressed states population in both closed-loop for $\varphi_0=0$ (a), $\pi$ (b) and open-loop (c) configurations. The parameters are same as in Fig. 2.

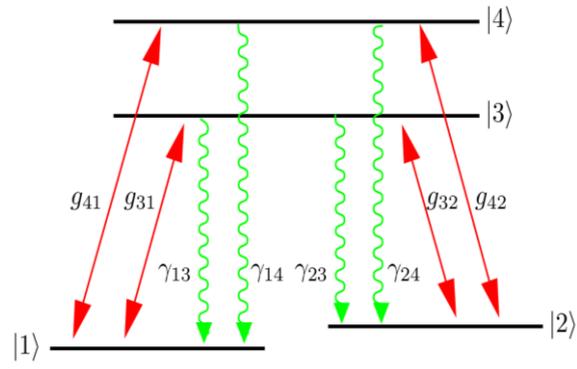

Figure 1

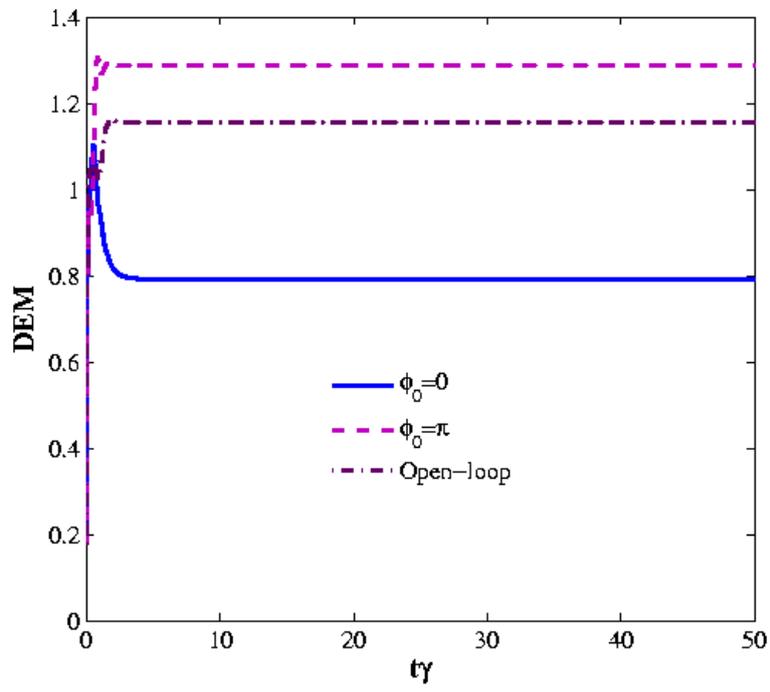

Figure 2

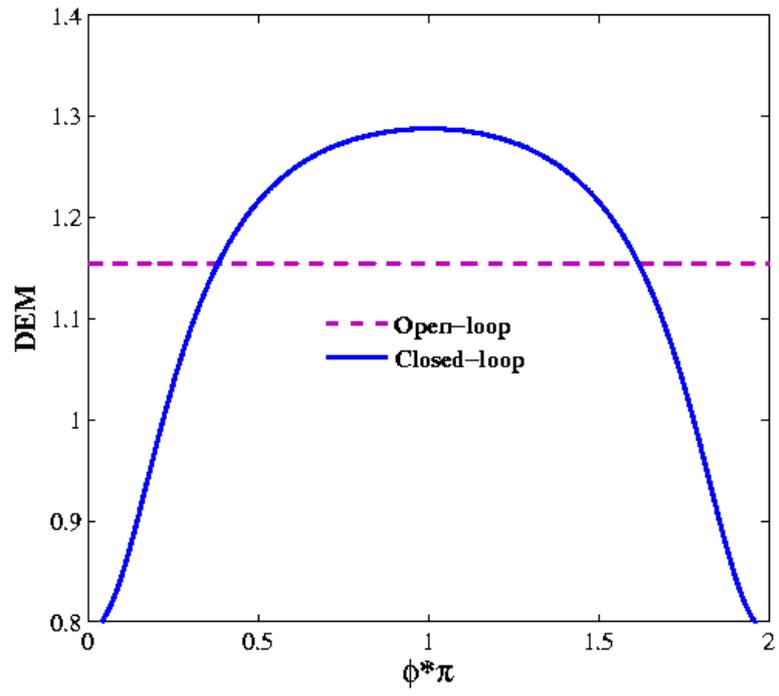

Figure3

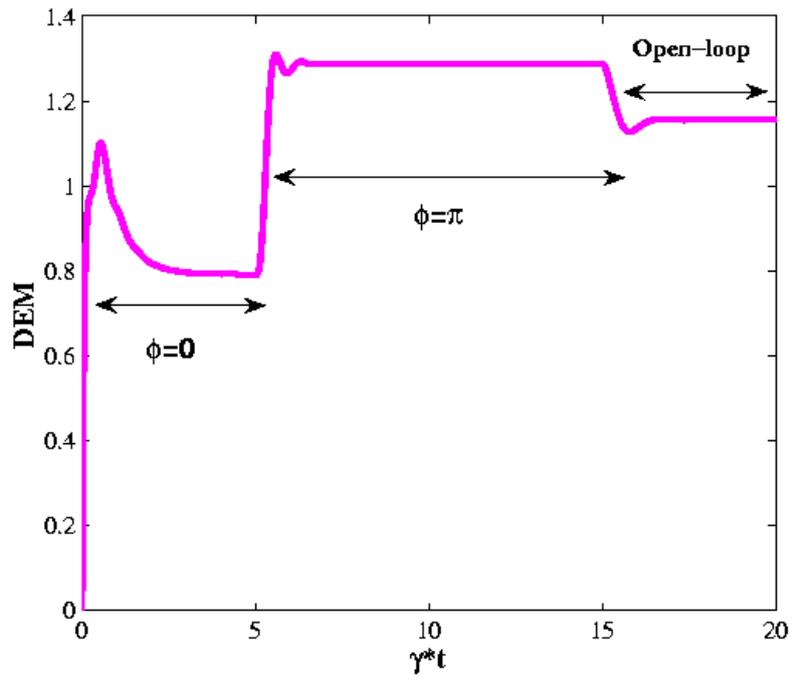

Figure 4

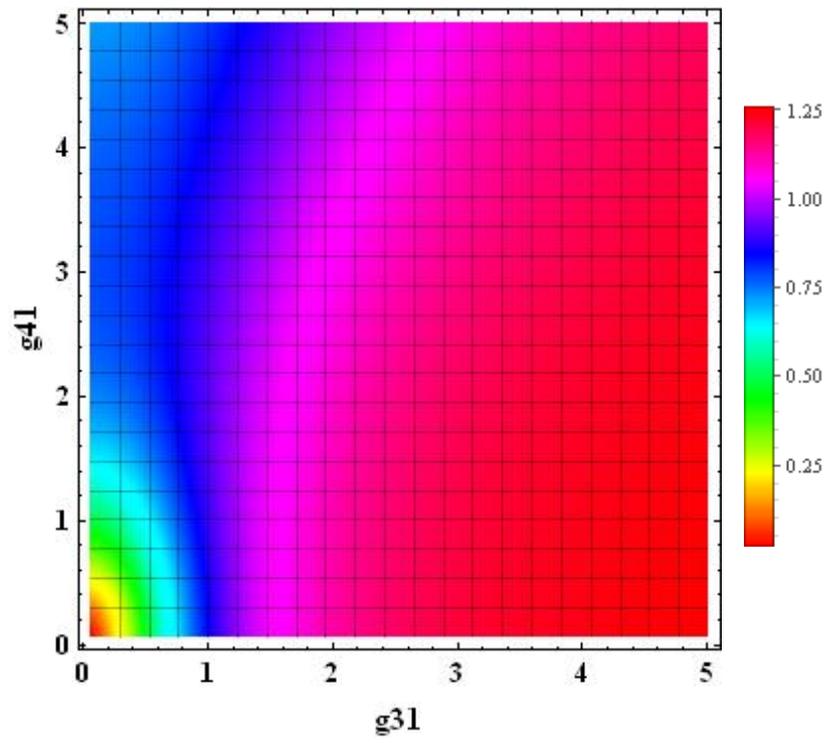

Figure 5

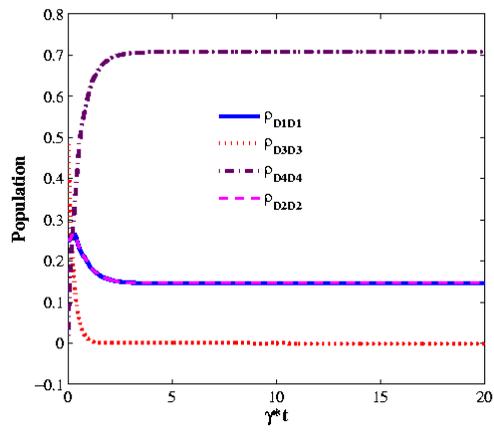

(a)

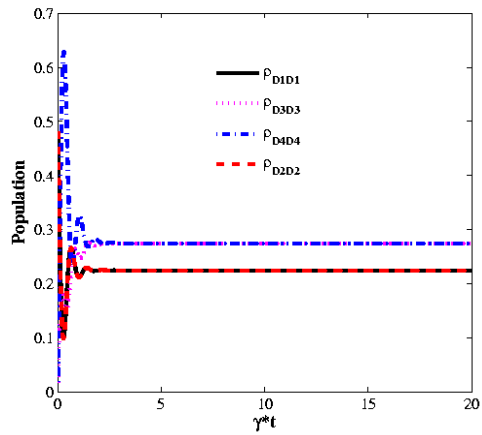

(b)

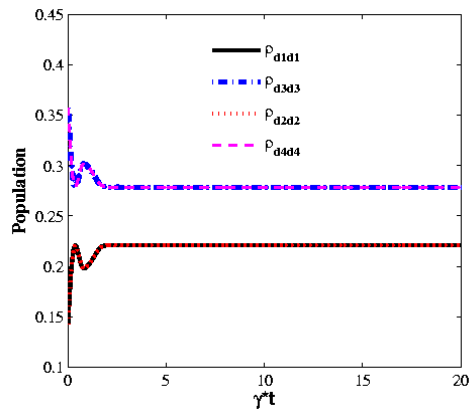

(c)

Figure 6